\documentclass[12pt]{article}

\usepackage{amssymb,amscd,amsmath}
\numberwithin{equation}{section}
\newcommand\rmd{{\rm d}}
\newcommand\rme{{\rm e}}

\begin{document}

\title{Totally Asymmetric Torsion on Riemann-Cartan Manifold}
\author{Yuyiu Lam\\
{\small Physics Department, Jinan University,
Guangzhou, 510632 China}\\
{\small Email: {\tt dr.lam@physics.org}}}

\date{\small November 1, 2002}
\maketitle

\begin{abstract}
A relativistic theory constructed on Riemann-Cartan manifold with a {\it derived}
totally antisymmetric torsion is proposed. It follows
the coincidence of the autoparallel curve and metric geodesic. 
The totally antisymmetric torsion $S_{[\lambda\mu\nu]}$
naturally appears in the theory without any {\it ad hoc} imposed on.
\end{abstract}

\section{Introduction}

Einstein-Cartan (EC) theory is one of the generalized Einstein's
General Relativity (GR) theories with Cartan's affine connection. The antisymmetric part
of affine connection namely torsion is introduced. Torsion
is suggested to act as a source for gravitational field (Hehl {\it etal} 1976 and de Sabbata {\it etal} 1985), and could be an essential
factor of quantization of spacetime (de Sabbata 1994) due to its
interpretation of topological defect of spacetime manifold. However, 
at first glance the autoparallel curve
\begin{equation}
\frac{\rmd^2x^\lambda}{\rmd s^2}+\Gamma^\lambda{}_{\mu\nu}
\frac{\rmd x^\mu}{\rmd s}\frac{\rmd x^\nu}{\rmd s}=0\label{1}
\end{equation}
with the presence of torsion and contorsion in the EC-theory does not seem to
coincide to the metric geodesic
\begin{equation}
\frac{\rmd^2x^\lambda}{\rmd s^2}+\{{}_\mu{}^\lambda{}_\nu\}
\frac{\rmd x^\mu}{\rmd s}\frac{\rmd x^\nu}{\rmd s}=0\label{2}
\end{equation}
where
\begin{equation}
\Gamma^\lambda{}_{\mu\nu}=\{{}_\mu{}^\lambda{}_\nu\}
+S_{\mu\nu}{}{}^\lambda-(S^\lambda{}_{\mu\nu}
+S^\lambda{}_{\nu\mu})\label{3}
\end{equation}
is the component of the affine connection which is generally not
symmetric, and $\{{}_\mu{}^\lambda{}_\nu\}$ is the component of the 
metric connection known as Christoffel symbol. This paradox
is so-called G-A problem which has recently been reviewed by Fiziev (1998).
Torsion with component
$S_{\mu\nu}{}{}^\lambda$ is usually denoted by the antisymmetric part of
the affine connection 
\begin{equation}
S_{\mu\nu}{}{}^\lambda=\frac{1}{2}(\Gamma^\lambda{}_{\mu\nu}
-\Gamma^\lambda{}_{\nu\mu}).\label{3a}
\end{equation}
or represented by torsion one-form. There would be many theoretical interesting
phenomena if torsion happens to be totally antisymmetric as shown by de
Sabbata and Gasperini (1985) and more recently developed by Hammond (1999). 
We attempt to show that torsion can reduce to
a totally antisymmetric tensor, resulting the coincidence of the affine and metric
geodesics. The totally antisymmetric torsion turns out to be a vector field or
scalar field which appears on the RHS of the proposed field equation similar
to many scalar-tensor theories as shown in the following sections.

\section{The affine connection}

We construct the theory on $U_4$ manifold as in EC-theory
with the metricity condition
\[
\nabla {\bf g}=0
\]
and is assumed to satisfy the principle of equivalence. It should be noted
that the $\nabla$ here is defined in terms of the affine connection
which is similar to EC-theory (Hehl {\it etal} 1976), related to the
metric connection by
\[
\Gamma^\lambda{}_{\mu\nu}=\{{}_\mu{}^\lambda{}_\nu\}
+S_{\mu\nu}{}{}^\lambda-(S^\lambda{}_{\mu\nu}+S^\lambda{}_{\nu\mu})
\]
as stated in \eqref{3}. At this moment we have to state a very useful theorem
here which appears in every differential geometry books:

\medskip\noindent
{\bf Theorem}\quad The exponential mapping $\exp_p$ is a diffeomorphism from a
neighborhood of $0\in T_p(M)$ to a neighborhood of $p\in M$.
\medskip

According to Straumann's analysis (Straumann 1984), if we choose a basis 
$e_1,\,\ldots,\,e_n$ of $T_p(M)$, then we can represent a neighborhood of $p$
uniquely by $\exp_p(x^ie_i)$. The set $(x^1,\,\ldots,\,x^n)$ are known as
{\it normal} or {\it Gaussian} coordinates. Since $\exp_p(sv)=\gamma_v(s)$
(for some neigbhorhood $v\in V$ of $0\in T_p(M)$, $\dot\gamma$ is autoparallel
along curve $\gamma$ and $s\in{\rm R}$), the curve $\gamma_v(s)$ has normal coordinates
$x^i=v^it$, with $v=v^ie_i$. In terms of these coordinates, \eqref{1} becomes
\[
\Gamma^\lambda{}_{\mu\nu}\frac{\rmd x^\mu}{\rmd s}\frac{\rmd x^\nu}{\rmd s}=0
\]
in $U_4$ manifold. Hence we have
\begin{equation}
\Gamma^\lambda{}_{\mu\nu}(0)+\Gamma^\lambda{}_{\nu\mu}(0)=0. \label{4}
\end{equation}
If the connection
is {\it symmetric}, it follows that $\Gamma^\lambda{}_{\mu\nu}(0)=0$.

The metric connection $\{{}_\mu{}^\lambda{}{}_\nu\}$ of course fulfils 
this situation according to standard GR.
Vanishing of $\Gamma^\lambda{}_{(\mu\nu)}$ and
$\{{}_\mu{}^\lambda{}{}_\nu\}$ of \eqref{3} in normal coordinates directly implies
\begin{equation}
(S^\lambda{}_{\mu\nu}+S^\lambda{}_{\nu\mu})_{p=0}=0.\label{5}
\end{equation}
This tensor relation is valid at all points
$p\in U\subset U_4$. It implies
$S^\lambda{}_{\mu\nu}=S^\lambda{}_{[\mu\nu]}$. Hence, with the definition of
torsion \eqref{3a} we have
\begin{equation}
S_{\lambda\mu\nu}=S_{[\lambda\mu\nu]}\label{6}.
\end{equation}
It follows that torsion in fact is a totally antisymmetric tensor as a
consequence of vanishing symmetric affine connection in local inertial
frame (Einstein elevator). This relation was firstly
proposed by Yu (1989), and the simplified torsion can reduce to an
axial vector or a scalar field if appropriate physical conditions
are imposed. Recently, this equation becomes the basic assumption for
Hammond to construct his relativistic theory to relate the current issues of
quantum gravity (Hammond 1999, Gruver {\it etal} 2001). However, the originality
of this complete antisymmetric property does not seem to be shown explicitly.

Equation \eqref{5} implies that the symmetric and antisymmetric parts of the 
affine
connection simply are related by
\[
\Gamma^\lambda{}_{\mu\nu}=\{{}_\mu{}^\lambda{}{}_\nu\}
+S_{\mu\nu}^{}{}^\lambda.
\]
It follows the equality of symmetric affine connection and metric 
connection, resulting the coincidence of the affine geodesic \eqref{1} and metric geodesic \eqref{2}
in our approach. This naturally comes from \eqref{4}, giving
a simpler geometry than that original torsion-contorsion picture in EC-theory,
with physical observational phenomenon.

\section{The field equations}

Having settled the relation \eqref{6}, the derivation of the field equations
are pretty straight forward.
The derivation of the field equations can be started from Cartan's
structure equations
\[
\Theta^\mu=\rmd\theta^\mu+\omega^\mu{}_\nu\wedge\theta^\nu
\]
\[
\Omega^\mu{}_\nu=\rmd\omega^\mu{}_\nu
+\omega^\mu{}_\lambda\wedge\omega^\lambda{}_\nu
\]
where $\Theta^\mu$ and $\Omega^\mu{}_\nu$ are the usual torsion one-form
and curvature two-form, with the spin connection $\omega^\mu{}_\nu$ and
anholonomic basis one-form $\theta^\mu$. And
\begin{equation}
\Omega^\mu{}_\nu=\frac{1}{2}R^\mu{}_{\nu\lambda\gamma}(\Gamma)
\theta^\lambda\wedge\theta^\gamma \qquad \Theta^\lambda=
\frac{1}{2}S_{\mu\nu}{}{}^\lambda\theta^\mu\wedge\theta^\nu\label{5a}
\end{equation}
are defined in terms of the components of Riemann tensor
$R^\mu{}_{\nu\lambda\gamma}(\Gamma)$ and torsion tensor
$S_{\mu\nu}{}{}^\lambda$ in orthonormal frame.

The component of Ricci tensor due to the contraction of the first and 
third indices is defined by
\[
R_{\mu\nu}(\Gamma)=R_{\mu\nu}-\nabla_\lambda S_{\nu\mu}{}{}^\lambda
-S_{\lambda\mu}{}{}^\rho S_{\nu\rho}{}{}^\lambda
\]
where $\nabla$ is the covariant derivative defined in terms of the 
Christoffel symbol. From now on, all the $R^\lambda{}_{\gamma\mu\nu}$, 
$R_{\mu\nu}$, $R$ and $G_{\mu\nu}$ etc. are defined in terms of the Christofel
symbols as in GR unless those tensors are specified by $\Gamma$.
Hence, we have a set of symmetric and antisymmetric
Ricci tensors, with components (we interchanged the indices of the
torsions, so they appear in positive signs here)
\begin{equation}
R_{(\mu\nu)}(\Gamma)=R_{\mu\nu}+S_{\mu\lambda}{}{}^\rho
S_{\nu\rho}{}{}^\lambda\label{5b}
\end{equation}
\begin{equation}
R_{[\mu\nu]}(\Gamma)=\nabla_\lambda S_{\mu\nu}{}{}^\lambda\label{5c}
\end{equation}

The component of the totally antisymmetric torsion $S_{[\lambda\mu\nu]}$ 
in \eqref{6} implies the existence of a 4-vector $P^\lambda$ such that
\begin{equation}
S_{\lambda\mu\nu}=\varepsilon_{\lambda\mu\nu\rho}P^\rho \qquad
S^{\lambda\mu\nu}=\varepsilon^{\lambda\mu\nu\rho}P_\rho \label{7}
\end{equation}
where $\varepsilon_{\lambda\mu\nu\rho}$ is an unit alternating tensor, 
the covariant derivative of it is zero (Schouten 1954)
\[
\nabla_\alpha\varepsilon_{\lambda\mu\nu\rho}=0.
\]

Substituting \eqref{7} into \eqref{5b}, having simplified gives
\begin{equation}
R_{(\mu\nu)}(\Gamma)=R_{\mu\nu}+2g_{\mu\nu}P_\alpha P^\alpha-2P_\mu
P_\nu\label{7a}
\end{equation}
and the required Ricci scalar by contraction is
\begin{equation}
R(\Gamma)=R+6P_\alpha P^\alpha.\label{7b}
\end{equation}

The divergence free field equation cannot simply be made by adding these 
two terms together with a $-\frac{1}{2}g_{\mu\nu}$ factor of \eqref{7b},
since the second contracted Bianchi identity of the generalized Einstein 
tensor does not vanish. Our approach is different from Yu's
derivation of the field equation (Yu 1989), he first proposed the existence
of the ``generalized full field equation'' such as
\begin{equation*}
R_{\mu\nu}(\Gamma)-\frac{1}{2}g_{\mu\nu}R(\Gamma)=\frac{8\pi G}{c^4}T_{\mu\nu}
\end{equation*}
where $T_{\mu\nu}$ is defined as the generalized energy momentum tensor with the
antisymmetric matter. Afterwards, he imposed the condition \eqref{6} on this to
simplify the suggested field equation resulting a different scalar field terms from ours
(the sign and coefficient of the scalar field terms) (see \eqref{14}).
Here, we suggest that the field equation
should be derived from the
variational principle with the Ricci scalar \eqref{7b} for a clear picture.

Before doing this let us first consider the antisymmetric Ricci tensor 
\eqref{5c} with the assumption relating to the antisymmetric matter
\begin{equation}
R_{[\mu\nu]}(\Gamma)=\nabla_\lambda S_{\mu\nu}{}{}^\lambda=\chi
T_{[\mu\nu]}
\label{11}
\end{equation}
where $T_{[\mu\nu]}$ is suggested to be interpretated as the energy momentum tensor
of intrinsic angular momentum of matter, or the intrinsic spin 
(Hehl {\it etal} 1976 and de Sabbata {\it etal} 1985), and $\chi$ is the
corresponding coupling constant. According to Eddington's principle of 
identification in GR (Eddington 1924), $T_{[\mu\nu]}$
should be related to both of the orbital angular momentum and spin of 
matter $M_{\mu\nu}{}{}^\lambda$, gives the conservation law for the total
angular momentum of field and matter via:
\begin{equation}
\nabla_\lambda(S_{\mu\nu}{}{}^\lambda+M_{\mu\nu}{}{}^\lambda)=0.\label{12}
\end{equation}
When outside the source i.e. $M_{\mu\nu}{}{}^\lambda=0$, torsion does not vanish in
general according to this equation. This may be interpreted as the 
inherent intrinsic spin of spacetime structure.

Using condition \eqref{7} again, \eqref{11} gives the antisymmetric field
equation in terms of the vector field $h_\alpha$
\begin{equation}
\partial_\mu P_\nu-\partial_\nu P_\mu=-\frac{\chi}{2}
\varepsilon_{\mu\nu\alpha\beta}T^{[\alpha\beta]}.\label{13}
\end{equation}
This equation can be rewritten in terms of the dual of antisymmetric Ricci
tensor $\tilde R_{[\mu\nu]}$ where $P^\lambda$ looks like the gauge potential
in electromagnetic theory
\[
\partial_\mu P_\nu-\partial_\nu P_\mu=\tilde R_{[\mu\nu]}
\]
and $\tilde R_{[\mu\nu]}$ becomes the field strength.

If there is no antisymmetric matter $T_{[\mu\nu]}=0$, \eqref{13}
gives two solutions
\begin{equation}
P_\nu=0 \qquad \mbox{or} \qquad P_\nu=\partial_\nu\phi \label{13-a}
\end{equation}
in a star-shaped region, where $\phi$ is a scalar (field).

Going back to the construction of (symmetric) field equation, substituting 
$P_\nu=0$ into \eqref{7b} gives the usual Ricci scalar in
GR. For $P_\nu=\partial_\nu\phi$, the Ricci scalar can be
expressed in terms of the scalar field
\begin{equation}
R(\Gamma)=R+6\partial_\alpha\phi\partial^\alpha\phi.\label{13.x}
\end{equation}

According to Tupper (1974), the variational principle
in such form
\begin{equation}
\delta\int_D(R+\lambda\partial_\alpha\phi\partial^\alpha\phi+L_M)
\sqrt{-g}\,\rmd^4x=0\label{13.y}
\end{equation}
always gives the field equation
\begin{equation}
G_{\mu\nu}=-\lambda\left(\partial_\mu\phi\partial_\nu\phi-\frac{1}{2}
g_{\mu\nu}\partial_\alpha\phi\partial^\alpha\phi\right)+\kappa T_{\mu\nu}
\label{13c}
\end{equation}
with the requirement of
\begin{equation}
\nabla_\alpha\partial^\alpha\phi=0\label{13b}
\end{equation}
where $\lambda$ is any real number, $\kappa=8\pi G/c^4$, $G_{\mu\nu}$ is
the usual Einstein tensor defined in terms of the Christofel symbol and 
$T_{\mu\nu}$ is the energy momentum tensor of matter excluding
the gravitational field. The derivation of the vacuum field equation is shown
in Appendix. Tupper (1974) and many others have shown that any
relativistic field
equations in this form always predict the three classical tests of GR.
It is indicated that the Lagrangian of matter $L_M$
includes all physical field except gravity. The assumption that $\phi$
satisfies \eqref{13b} in order to the paths of test particles be
geodesics implies that the scalar field {\bf does not} couple with (symmetric) matter.
Actually it has been demonstrated for the coincidence of the affine and 
metric geodesics in the previous section. By varying the Lagrangian $L_M$ in
\eqref{13.y} with respect to
$g_{\mu\nu}$, we have the usual energy momentum tensor of matter in
\eqref{13c}.

Putting $\lambda=6$, we obtain the full field equation
\begin{equation}
G_{\mu\nu}=-6\left(\partial_\mu\phi\partial_\nu\phi-\frac{1}{2}g_{\mu\nu}
\partial_\alpha\phi\partial^\alpha\phi\right)+\kappa T_{\mu\nu}\label{14}
\end{equation}
for outside or non-existence of antisymmetric matter $T_{[\mu\nu]}$.
Nevertheless, we still have to justify the requirement \eqref{13b}. Taking
the second contracted Bianchi identity $\nabla_\nu G^{\mu\nu}=0$, 
\eqref{14} becomes
\[
(\nabla_\nu\partial^\nu\phi)\partial^\mu\phi+\kappa\nabla_\nu
T^{\mu\nu}=0.
\]
For satisfying the covariant conservation law of matter and field, i.e.
$\nabla_\nu T^{\mu\nu}=0$, implies
\[
(\nabla_\nu\partial^\nu\phi)\partial^\mu\phi=0
\]
giving the non-trivial result \eqref{13b} as $\partial^\mu\phi\neq 0$.

The extra scalar field terms in most scalar-tensor theory or EC-theory might
be interpreted as the ``energy momentum of gravitational field''. However, the 
scalar terms in \eqref{14} seem to have counter-contribution of gravity, it contradicts
to our common sense that gravity should be universal attractive. After all we do
not find any anti-gravity matter. It should be noticed that the scalar field here
does not come from (symmetric) matter, which is simply derived from the Rieman-Cartan
geometry from the LHS. Or more appropriate it could be interpreted as
the consequence of the inherent spacetime fabric translation due to torsion (Parallelly
transporting a vector around a closed loop results the misfit to the original vector. The
rotation of the transformed vector reveals curvature and the displacement reveals torsion).
We might think such topological defect acts like a repulsive gravity at the very
structure of spacetime, ie. the quantum nature of spacetime. Having this idea
singularities might be removed here, and it might act as a low energy limit
cut-off in the (super)string and supergravity theory.

\section{Conformal to Brans-Dicke theory}
The vacuum field equation of \eqref{14} can be constructed from
\begin{equation}
R_{\mu\nu}=-\lambda\partial_\mu\phi\partial_\nu\phi\label{15}
\end{equation}
where it is the usual Ricci tensor defined in terms of the Christoffel 
symbols, contrary to the generalized Ricci tensor \eqref{7a}.

Consider two conformally related spacetime manifolds $\bar M$ and $M$
such that their metric satisfy
\[
g_{\mu\nu}=\rme^{2\phi}\bar g_{\mu\nu}.
\]
The components of the Ricci tensors of $\bar M$ and $M$ are related by 
(Einsenhart 1925)
\begin{equation}
R_{\mu\nu}=\bar R_{\mu\nu}+2\nabla_\mu\partial_\nu\phi-2\partial_\mu\phi
\partial_\nu\phi+\bar g_{\mu\nu}\nabla_\alpha\partial^\alpha\phi
+2\bar g_{\mu\nu}\partial_\alpha\phi\partial^\alpha\label{16}
\end{equation}
where $\phi$ denotes the scalar field in \eqref{15}, $R_{\mu\nu}$ and
$\bar R_{\mu\nu}$ are the corresponding Ricci tensors in the spacetime 
manifolds $M$ and $\bar M$, respectively.

Substituting \eqref{15} into \eqref{16} we have the field equation in the
spacetime manifold $\bar M$
\begin{equation}
\bar R_{\mu\nu}+2\nabla_\mu\partial_\nu\phi+(\lambda-2)\partial_\mu\phi
\partial_\nu\phi+\bar g_{\mu\nu}\nabla_\alpha\partial^\alpha\phi
+2\bar g_{\mu\nu}\partial_\alpha\phi\partial^\alpha\phi=0.\label{17}
\end{equation}
If we put
\[
\phi=\frac{1}{2}\ln\varphi
\]
\eqref{17} becomes
\begin{equation}
\bar R_{\mu\nu}+\frac{1}{\varphi}(\nabla_\mu\partial_\nu\varphi
+\frac{1}{2}\bar g_{\mu\nu}\partial_\alpha\varphi\partial^\alpha\varphi)
+\frac{1}{4}(\lambda-6)\frac{1}{\varphi^2}\partial_\mu\varphi\partial_\nu
\varphi=0.\label{18}
\end{equation}
This equation is identical with the vacuum field equation of Brans-Dicke
theory with the identification
\[
\lambda=4\omega+6
\]
where $\omega$ is the Brans-Dicke parameter.

It can be shown that if $\nabla_\alpha\partial^\alpha\phi=0$ in $M$, the
scalar field satisfies
\[
\nabla_\alpha\partial^\alpha\varphi=0
\]
in the spacetime manifold $\bar M$.

It is clearly seen that the proposed theory with $\lambda=6$ is
conformally related to the Brans-Dicke theory in vacuum. This leads to a
vanishing Brans-Dicke parameter $\omega$ for the conformal relation.

\section{Conclusion}

The construction and derivation of this theory is nothing new which is just
based on the Riemann-Cartan spacetime manifold, however, with the importance
of the vanishing affine connection property in local inertia frame. Nevertheless, it is not
a special case of EC-theory with a totally antisymmetric
torsion tensor where the vector field (or scalar field) interacts to the
Lagrangian of matter $L_M$. The derived scalar field $\phi$ here does not
couple to matter field in the variational principle \eqref{13.y}. This
assumption is due to a test particle always moving along a geodesic as stated before.

On the other hand, the scalar field $\phi$ originates from \eqref{13} when
$T^{[\mu\nu]}=0$. It is safe to put $T^{[\mu\nu]}$ equal to zero
even the suggestion of the relationship between antisymmetric matter and
intrinsic spin is controversial. Down to the worst case, we still have one of the
solution in \eqref{13-a} with $P^\lambda=0$, this gives the usual Einstein's GR.
It should be noted that torsion and contorsion need to vanish if EC-theory
reduces to GR. However, we do not pre-suppose the vanishing of them as mentioned 
above.

The extra scalar term with a negative sign in \eqref{14} has some interesting
features mentioned in section 3. We think that it is worth to further
investigate for quantum geometry.

\appendix
\setcounter{section}{1}
\section*{Appendix}
\subsection*{Hamilton's principle for the vacuum field equation}
The hamiltonian variational principle for the field equations in vacuum is
\begin{equation}
\delta\int_D R(\Gamma)\sqrt{-g}\,\rmd^4x=0\label{a1}
\end{equation}
where $R$ is the Ricci scalar given in \eqref{13.x}, $D$ is a compact
region having a smooth boundary $\partial D$. The variation of the metric
are assumed vanish on $\partial D$. \eqref{a1} becomes
\begin{equation}
\int_D\delta R_{\mu\nu}(\Gamma)g^{\mu\nu}\sqrt{-g}\,\rmd^4x+\int_D
R_{\mu\nu}(\Gamma)\delta(g^{\mu\nu}\sqrt{-g})\,\rmd^4x=0.\label{a3}
\end{equation}

The component of the Ricci tensor is defined by (from \eqref{7a})
\begin{equation}
R_{\mu\nu}(\Gamma)=R_{\mu\nu}+2(g_{\mu\nu}\partial_\lambda\phi
\partial^\lambda\phi-\partial_\mu\phi\partial_\nu\phi).\label{a4}
\end{equation}
Similar to GR, $g^{\mu\nu}\delta R_{\mu\nu}$ vanishes on
the boundary of $D$. Hence, from \eqref{a4}, after calculated we have
\begin{equation}
g^{\mu\nu}\delta R_{\mu\nu}(\Gamma)=2g^{\mu\nu}\delta g_{\mu\nu}
\partial_\lambda\phi\partial^\lambda\phi+8\delta g^{\mu\nu}\partial_\mu
\phi\partial_\nu\phi+6g^{\mu\nu}\delta(\partial_\mu\phi\partial_\nu\phi).
\label{a5}
\end{equation}
However, the last term of above equation can be written in terms of the
divergence
\[
6g^{\mu\nu}\delta(\partial_\mu\phi\partial_\nu\phi)=6g^{\mu\nu}\delta
\partial_\mu(\phi\partial_\nu\phi)
\]
where an assumed condition $\partial_\lambda\partial^\lambda\phi=0$ is
imposed. This vanishes on the boundary of $D$ by Gauss' theorem. Hence
\eqref{a5} is simplied as
\begin{equation}
g^{\mu\nu}\delta R_{\mu\nu}(\Gamma)=2g^{\mu\nu}\delta g_{\mu\nu}
\partial_\lambda\phi\partial^\lambda\phi+8\delta g^{\mu\nu}\partial_\mu
\phi\partial_\nu\phi.\label{a10}
\end{equation}

Using the relations
\[
\delta g^{\mu\nu}=-g^{\mu\alpha}g^{\nu\beta}\delta g_{\alpha\beta}
\]
\[
\delta(g^{\mu\nu}\sqrt{-g})=\sqrt{-g}\left(\frac{1}{2}g^{\mu\nu}
g^{\alpha\beta}-g^{\mu\alpha}g^{\nu\beta}\right)\delta g_{\alpha\beta}
\]
and substituting \eqref{a4} and \eqref{a10} into \eqref{a3}, having simplied
we obtain
\begin{eqnarray*}
&\quad\delta\int_D R(\Gamma)\sqrt{-g}\,\rmd^4x\\
 &=-\int_D\left(R^{\mu\nu}-\frac{1}{2}g^{\mu\nu}R\right)\delta g_{\mu\nu}
\sqrt{-g}\,\rmd^4x\\
  &\quad+ \int_D2(g_{\mu\nu}\partial_\lambda\phi\partial^\lambda\phi-
\partial_\mu\phi\partial_\nu\phi)\left(\frac{1}{2}g^{\mu\nu}
g^{\alpha\beta}-g^{\mu\alpha}g^{\nu\beta}\right)\delta g_{\alpha\beta}
\sqrt{-g}\,\rmd^4x\\
  &\quad +\int_D(2g^{\alpha\beta}\delta g_{\alpha\beta}\partial_\lambda
\phi\partial^\lambda\phi-8\delta g_{\alpha\beta}\partial^\alpha\phi
\partial^\beta\phi)\sqrt{-g}\,\rmd^4x\\
&=-\int_D\left[G^{\mu\nu}+6\left(\partial^\mu\phi\partial^\nu\phi-
\frac{1}{2}g^{\mu\nu}\partial_\alpha\phi\partial^\alpha\phi\right)\right]
\delta g_{\mu\nu}\sqrt{-g}\,\rmd^4x.
\end{eqnarray*}
It follows the vacuum field equation in covariant form
\[
G_{\mu\nu}=-6\left(\partial_\mu\phi\partial_\nu\phi-\frac{1}{2}g_{\mu\nu}
\partial_\alpha\phi\partial^\alpha\phi\right)
\]
where $G_{\mu\nu}$ is the component of the usual Einstein tensor defined
in terms of the Christoffel symbol.

\end{document}